\documentclass[aip,pof,amsmath,amssymb,reprint]{revtex4-1} \synctex=1 

\usepackage{graphicx}
\usepackage[colorlinks=true,linkcolor=blue]{hyperref}
\usepackage{hyperref}
\usepackage{bm}

\begin{document}


\title {Slip behavior during pressure driven flow of Laponite
  suspension}

\author{Prophesar M. Kamdi} 

\author{Ashish V. Orpe} \email{av.orpe@ncl.res.in} 

\affiliation{CSIR-National Chemical Laboratory, Pune 411008 India}
\affiliation{Academy of Scientific and Innovative Research (AcSIR), Ghaziabad 201002 India}

\author{Guruswamy Kumaraswamy}
\altaffiliation{Currently at Indian Institute of Technology Bombay, Powai, Mumbai 400076 India} 
\email{guruswamy@iitb.ac.in}

\affiliation{CSIR-National Chemical Laboratory, Pune 411008 India}

\begin{abstract}
  We investigate pressure driven pipe flow of Laponite suspension, as
  a model thixotropic fluid.  The tendency of the suspension to age is
  controlled by addition of sodium chloride salt to vary the ionic
  strength. We use a syringe pump to prescribe the flow and observe
  that a steady state flow is obtained.  Unusually, the steady state
  pressure drop required to maintain a constant flow rate decreases
  with increase in flow rate, in qualitative contrast to the
  expectation for Poiseuille flow. We demonstrate that experimental
  results obtained by varying the flow rate, salt concentration and
  flow geometry (pipe diameter and length) can be collapsed onto a
  single universal curve, that can be rationalized by invoking slip of
  the suspension at the tube walls. The Laponite suspension exhibits
  plug-like flow, yielding at the tube walls. Our results suggest that
  the slip length varies linearly with the flow rate and inversely
  with the tube diameter.
\end{abstract}

\maketitle

\section{Introduction}
The no-slip boundary condition is commonly invoked at liquid-solid
interfaces. This dictates that the liquid velocity at the interface is
equal to that of the solid. However, depending on the characteristics
of the liquid or that of the wall, liquids can slip at the interface.
For Newtonian liquids, slip has mostly been attributed to
non-wettability of the
surface~\cite{schnell56,watanabe99,tretheway02,li19,churaev84} with a
shear dependent slip length obtained theoretically for laminar
flows~\cite{aghdam16}. For a laminar flow flow through a pipe, slip
results in a distinct deviation from Poiseuille flow and, can get
further exaggerated with decrease in system size~\cite{cheng02}. An
elegant theoretical treatment by Lauga and Stone~\cite{lauga03}
accounts for slip during pressure driven flow through a capillary in
terms of an effective slip length that depends on system and material
properties. Others have applied a similar methodology to investigate
the slip during flow of visco-plastic fluids in various geometries,
and have demonstrated a scaling law for the slip velocity expressed in
terms of wall shear stress and slip
length~\cite{kalyon05,kalyon12}. The predictions of the velocity
profiles incorporating the scaling law, were shown to agree well with
experimental measurements~\cite{ortega16}. In a later experimental
study, it was shown that the scaling law holds only when the material
has yielded across the entire flowing cross-section but not for
co-existing solid-liquid regions or flows below the yield
threshold~\cite{poumaera14}. For shear-thinning visco-plastic fluids,
wall slip in the regime beyond yielding, was shown to be enhanced
compared to that for Newtonian liquids, while the opposite was
observed for shear-thickening fluids~\cite{haase17}.

A common feature running across all the studies described above is the
time independent nature of the fluid response. However, fluids that
exhibit time dependent or thixotropic behavior are not uncommon. For
example, crude oil that is pumped over large distances in pipelines
comprises a thixotropic paraffin gel~\cite{sastry11}. In the human
body, biliary sludge comprising bacteria and particulates flows
through viscoelastic gastrointestinal channels or through artificial
stents~\cite{berkel05, donelli07}.  The effect of slip on pressure
driven flow of thixotropic fluids remains poorly
understood~\cite{jamali17} which makes it that much difficult to
recover simple scaling laws characterising the slip behavior. A recent
theoretical work showed that while the steady state flow condition for
a thixotropic fluid is uniquely determined by the applied stress, the
evolution to steady state is determined by the build up and breakage
of the fluid microstructure~\cite{cunha20}.

Here, we experimentally study the pressure driven flow of a model
thixotropic fluid through a cylindrical tube to investigate the
influence of slip on overall flow. We employ aqueous suspensions of an
inorganic synthetic clay, Laponite, as a model thixotropic fluid,
whose time dependent response can be controlled by addition of
salt. Laponite comprises disk-shaped particles of $\approx 25$ nm
diameter and $0.92$ nm thickness.  When Laponite is vigorously stirred
into water, clay platelets disperse to form a clear suspension. On
ageing, platelet-platelet interactions result in the formation of
larger scale microstructure in the dispersion such that it becomes
increasingly viscous. At moderate concentrations, of the order of a
few percent (by weight), aqueous Laponite dispersions develop a yield
stress and exhibit solid-like response.  Addition of sodium chloride
salt to the aqueous dispersion modifies platelet-platelet
electrostatic interactions such that the ageing process is
significantly accelerated.  Time dependent structure formation and
relaxation phenomena in Laponite have been extensively researched in
the last few
decades~\cite{bonn99,knaebel00,bonn02,ruzicka10,ruzicka11,
  joshi14,joshi18a}. Researchers have reported a state diagram that
describes the non-equilibrium structure formation in Laponite as a
function of clay concentration and ionic
strength~\cite{ruzicka11}. The precise physical mechanism for
self-assembly in Laponite dispersions remains contentious, with
arguments for both
gel~\cite{bonn99,knaebel00,Levitz00,Bonn98,Tanaka04} and
glass~\cite{Mourchid95,Pignon96,Pignon97,PignonE97,Kroon96,Kroon98,Avery86,Nicolai00}
formation. Our work does not address the question of the underlying
structure that determines the rheology of Laponite dispersions. Of
interest to us is only that Laponite self assembles in water to form a
thixotropic dispersion, and that the paste-like rheology of Laponite
dispersions is strongly influenced by the addition of sodium chloride
salt.

\section{Experimental details}
\subsection{Materials}
Laponite was obtained from BYK additives Ltd. UK, and was used as
received. Sodium chloride was obtained from Merck Specialities
Pvt. Ltd. Mumbai and was used as received. Distilled deionized water
(conductivity = $18$ $\mu$S$/$cm) was obtained from
Millipore system (Merck Specialities Pvt. Ltd, Mumbai) and was used
immediately. 

\begin{figure*} \includegraphics[scale=0.5]{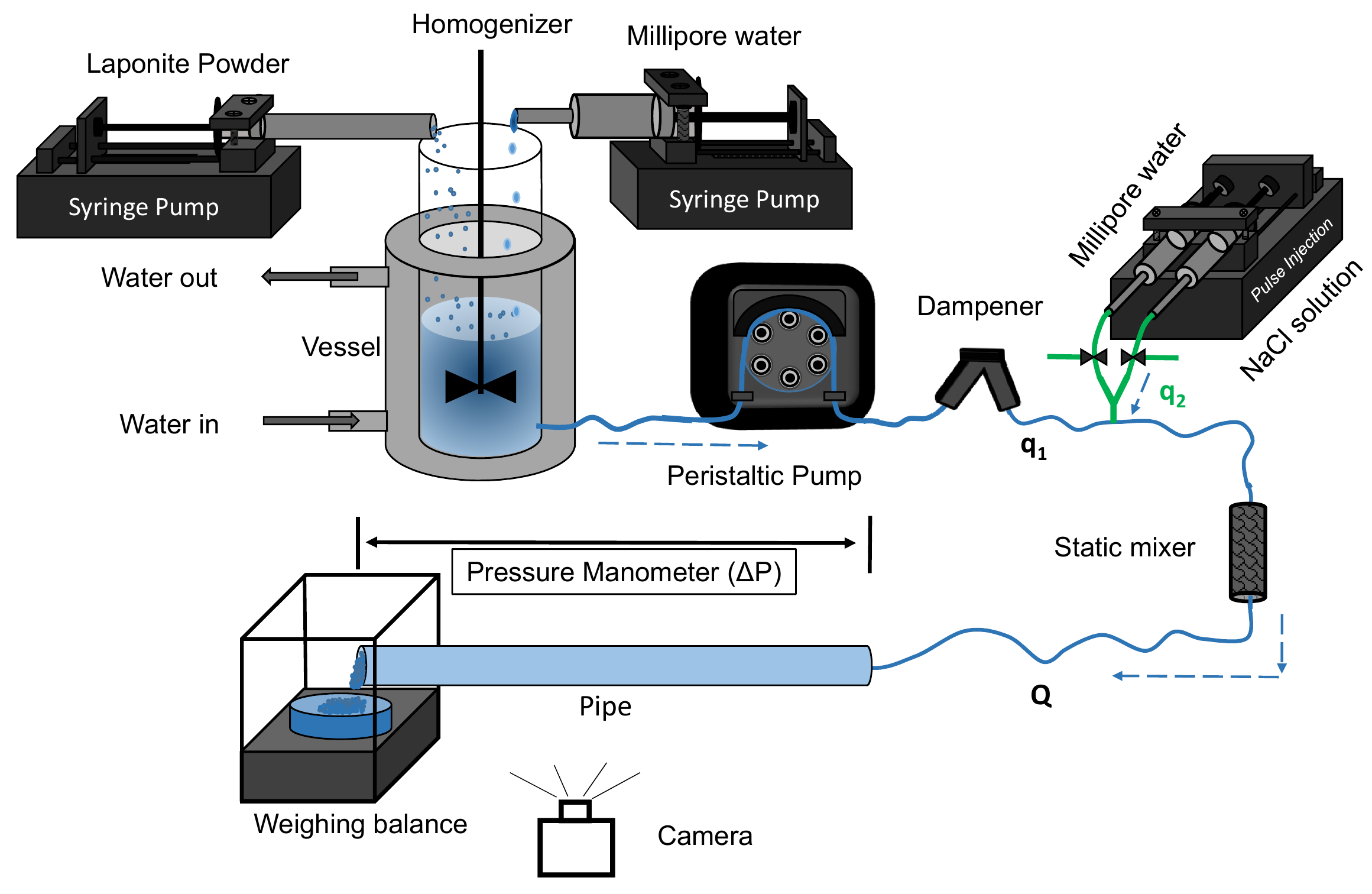}
  \caption{Schematic of the experimental assembly showing continuous
    preparation and pumping of Laponite suspension through a tube. The
    assembly allows addition of salt (NaCl) solution in fixed
    quantities and for a fixed duration.}
  \label{exp-schem}
\end{figure*}

\subsection{Methods}

We measured the pressure drop required to flow aqueous Laponite
dispersion at a constant flow rate through a tube of length, $L$, and
diameter, $D$, as shown in fig.~\ref{exp-schem}. Dry Laponite powder
and deionized water (Millipore) were continuously added to a vessel
and were subjected to high speed shearing (Ultraturrax homogenizer,
IKA). The vessel was kept immersed in a water bath to maintain the
temperature of Laponite-water mixture at $25^{0}$C throughout the
experiments. The outflow rate ($q_{1}$) from the vessel, as set by the
peristaltic pump was the same as the inlet flow rate to the vessel
which was set by the syringe pump. This ensured that volume of
suspension ($V$) in the vessel is maintained constant throughout. The
average residence time of the material in the mixer is maintained at
$45$ min. The high shear mixing protocol ensured that the state of the
Laponite dispersion remained constant throughout the duration of the
experiment. This was verified by measuring the outflow mass flow rate
from the vessel which remained constant in time.  To damp out
fluctuations due to the peristaltic pump, the suspension is passed
through a dampener (see section I in supplementary material).
Downstream of the dampener, we introduce a second stream (volumetric
flow rate, $q_2 = 0.13 q_1$) using a needle and mix these streams by
passing them through a pinched tube static mixer (see section II in
supplementary material). The syringe pump that injects the second
stream has two barrels so that it can flow either deionized water, or
a salt (NaCl) solution for a fixed duration. After mixing in the
static mixer, the suspension is passed through the tube.  The tube
entrance has a uniformly diverging taper to minimize entrance effects
due to change in cross-sectional area when the flow enters the tube.
The tube exit is open to atmosphere and we use a pressure gauge just
before the tube entrance to measure the pressure drop across the tube
length. The total volumetric flow rate ($Q = q_{1} + q_{2}$) in the
system is set by the peristaltic and syringe pumps.  The measured
outflow mass flow rate from the tube remained constant in
time. Experiments are performed for three flow rates ($Q = 0.46$,
$0.69$ and $0.92$ cm$^{3}$/min) and at three salt concentrations ($6$,
$9$ and $12$ mM, after the two streams are mixed). Tubes with
different dimensions were used in the experiments: Glass tube (lengths
$15.5$ and $30$ cm each with an internal diameter of $0.4$ cm and
$0.72$ cm) and steel tube (length $23$ and $53$ cm each with an
internal diameter of $0.1$ cm).  The Laponite concentration in all
experiments was kept unchanged throughout at $3.1$ wt \%.

\subsection{Pressure drop measurements}

\begin{figure} \includegraphics[scale=0.52]{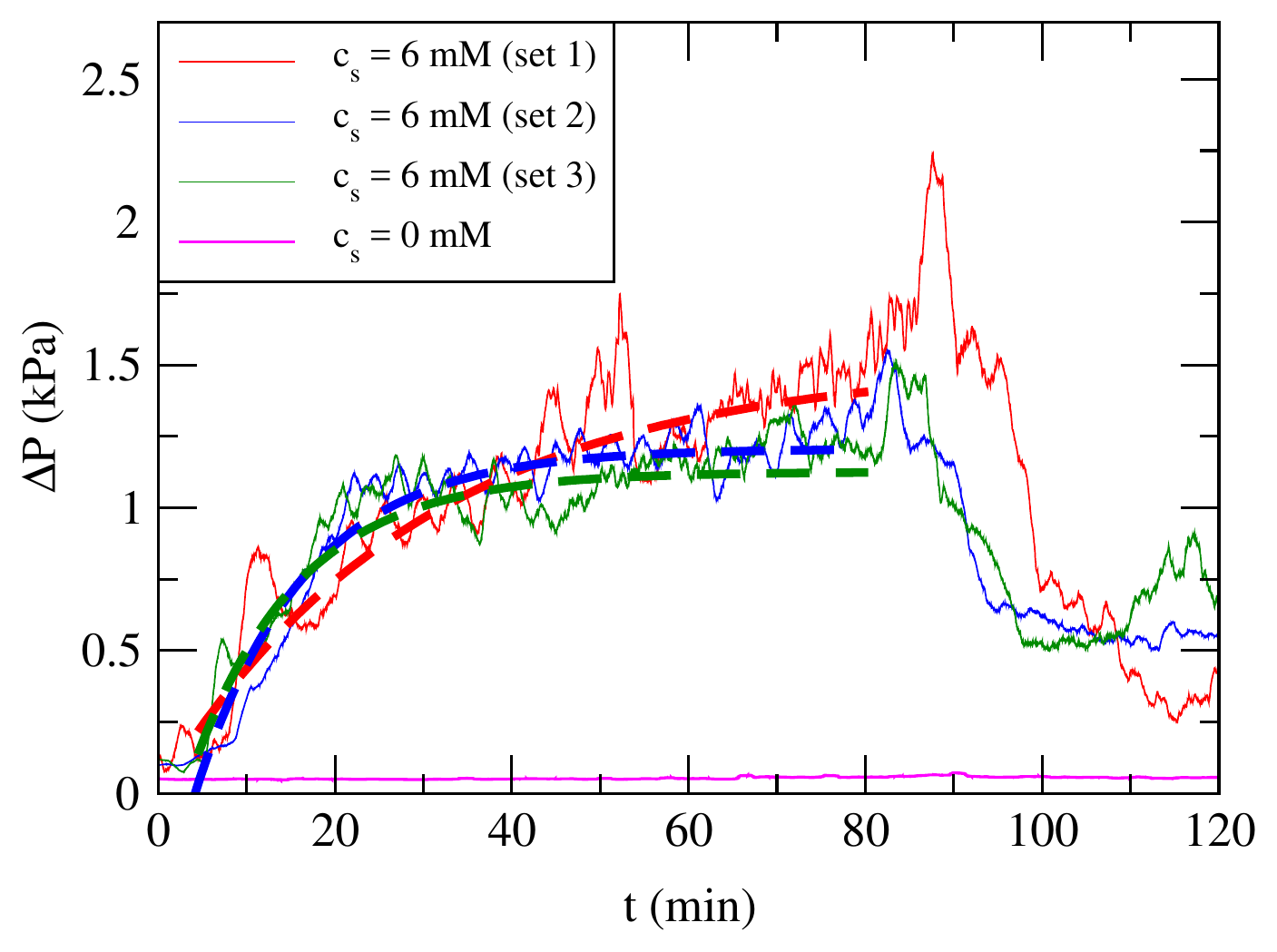}
  \caption{Time (t) dependent pressure drop ($\Delta P$) for the flow
    of Laponite suspension at a constant rate ($ Q = 0.46$
    cm$^{3}$/min) through the tube of length ($ L = 15.5$ cm) and
    diameter ($ D = 0.72$ cm) for a fixed salt concentration
    ($ C_{s} = 6$ mM).  Magenta colored solid line represents data
    without addition of salt, while three solid lines of different
    color (red, green and blue), respectively, represent three
    independent runs for addition of salt solution for a pulse of 80
    minutes duration ($ C_{s} = 6$ mM). The dashed lines of red, green
    and blue color represent a fitted non-linear equation to the
    respective pressure data to extract the saturated value of
    pressure drop $(\Delta P_{s}$. (see text for more details).}
  \label{s-pulse}
\end{figure}

The time variation of the measured pressure difference ($\Delta P$)
across the cylindrical tube during the flow of Laponite suspension
(with or without added salt), is shown in fig.~\ref{s-pulse} for a
fixed tube diameter, fixed salt concentration and fixed flow rate.
The variation of pressure difference for the Laponite suspension
without salt is quite negligible and remains constant throughout at
$0.06$ kPa. We refer to this value as the base or reference pressure.

The addition of salt solution pulsed for $\Delta t_{p}$ = 80 mins,
however, increases the pressure difference significantly above the
base value and achieves a near saturated state (with associated
fluctuations). These are shown for three independent runs (red, green
and blue solid lines in fig.~\ref{s-pulse}). Presence of high salt
concentration in Laponite suspension causes it to age rapidly with an
evolving microstructure comprising sample spanning aggregates, thereby
increasing the overall viscosity. This increase in the viscosity leads
to increased pressure difference above the base value. The near
saturated value of the pressure drop reflects a balance in the
microstructure of the Laponite suspension, between the salt-induced
ageing and flow effects. Following the stoppage of the salt solution
pulse after $\Delta t_{p}$, the pressure difference returns to the
base value over time suggesting that the salt has been removed from
the tube completely during that time. It can be noted that the
timescales leading to the saturated value of pressure drop and while
reverting to the base value are not the same. We believe that this
difference owes its origin to (i) significant heterogeneity in the
microstructures formed within the system and (ii) possibility of
traces of microstructures remaining in the tube over much longer time
duration while allowing for the flow of Laponite suspension (without
salt) associated with pressure drop larger than the base
value. Similar qualitative behavior for pressure drop variation is
also observed for different tube diameters, lengths, salt
concentration and flow rates investigated in this work.

The value of the plateau or the saturated pressure drop
($\Delta P_{s}$) is extracted by fitting the data for the 80 minutes
salt solution pulse with an exponential expression of the form
$ \Delta P = C_{1} (1 - exp(-t/\tau)) + C_{2}$. The saturation value
($\Delta P_{s}$) is then obtained as $ C_{1} + C_{2}$. The fit of this
equation to the three independent runs is shown as dashed lines in
red, green and blue colour in fig.~\ref{s-pulse}. The parameters
$C_{1}$, $C_{2}$ and $\tau$ were simply used to fit the expression to
the data and are observed to vary across three independent sets.  We note
similar time dependence for the pressure difference for all the flow
rates, tube diameters and salt concentrations employed. For each case,
$2$ or $3$ independent sets were performed and the data is reported
for all the sets in subsequent figures.

\section{Results and Discussion}

In Secs. IIIA and IIIB, we discuss the effect of tube length, flow
rate, tube diameter and salt concentration on the observed variations
in the saturated value of pressure drop ($\Delta P_{s}$). We, then,
try and show the scaling of all the data in Sec. IIIC based on simple flow modeling
and heuristic arguments.

\subsection{Effect of tube length and flow rate on saturated pressure drop}

\begin{figure} \includegraphics[scale=0.6]{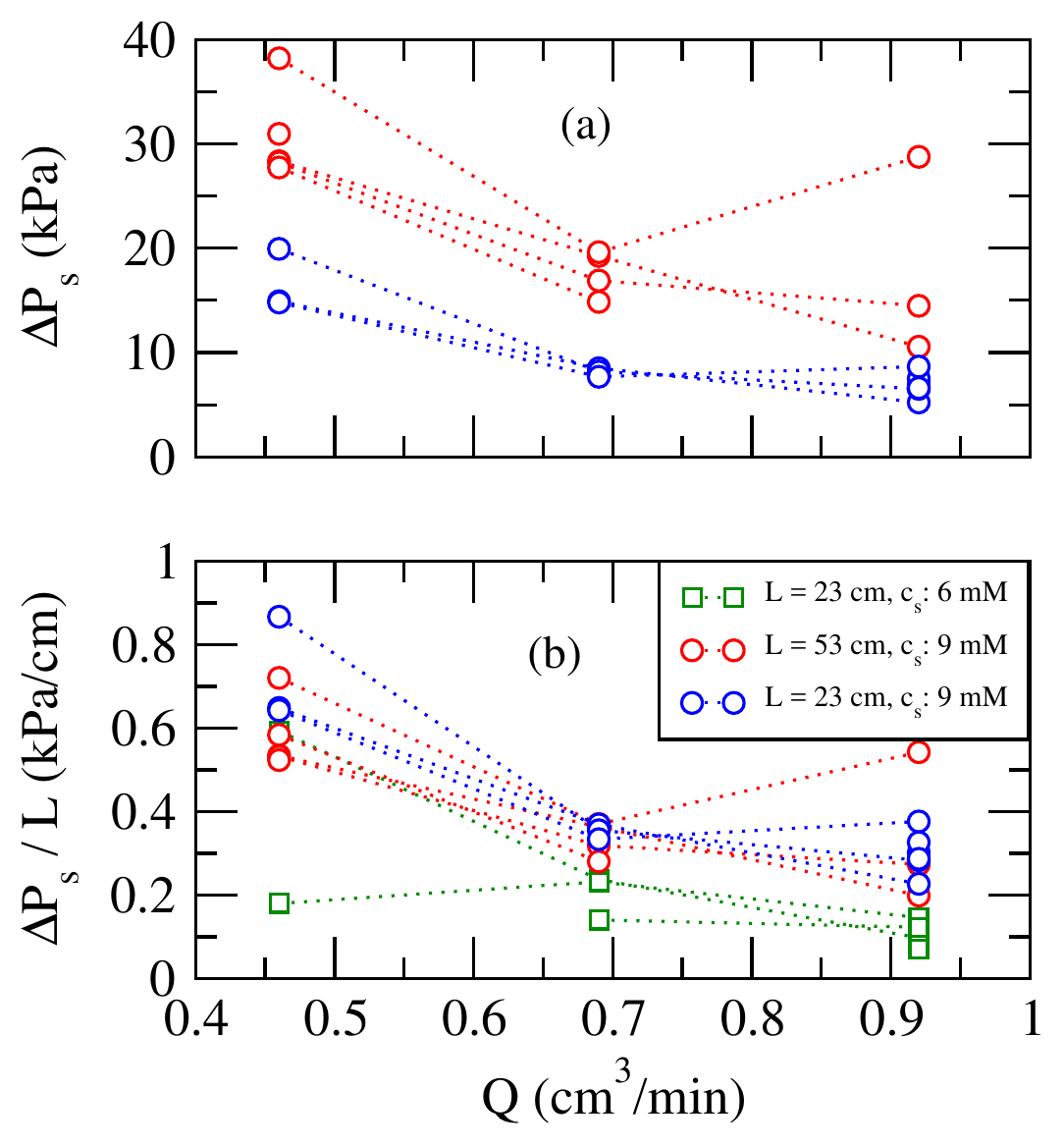}
  \caption{Variation of (a) saturated pressure drop ($\Delta P_{s}$)
    with flow rate in a tube of diameter ($D = 0.1$ cm) and salt
    concentration ($c_{s} = 9$ mM) for two different tube lengths and
    (b) Variation of $\Delta P_{s}/L$ with flow rate in a tube of
    diameter ($D = 0.1$ cm) and two salt concentrations and tube
    lengths.}
  \label{delp-lq}
\end{figure}

The variation of the saturated state pressure drop ($\Delta P_{s}$) with
flow rate ($Q$) for two different tube lengths, at fixed tube diameter
($D = 0.1$ cm) and salt concentration ($c_{s} = 9$ mM) is shown in
fig.~\ref{delp-lq}a. For a given flow rate, the pressure drop
increases with increase in the tube length. However, the pressure drop
per unit length ($\Delta P_{s}/L$) is nearly independent of tube
length as shown in fig.~\ref{delp-lq}b. This suggests that the
observed behavior and underlying mechanism is not localised, but it
remains the same everywhere along the tube length.

Interestingly, the data in fig.~\ref{delp-lq}b shows that the
saturated pressure drops decreases monotonically with increase in the
flow rate.  Further, this decrease in the pressure drop is more
prominent at higher salt concentration. At much higher salt
concentration ($c_{s} = 12$ mM), rapid increase in viscosity led to
intermittent flow causing problems in pressure drop measurements.
This behavior of pressure drop is qualitatively opposite to that
predicted by the Poiseuille equation. To rationalize this, we consider
the possibility that higher shear rates at higher $Q$ might result in
a flow-induced breakdown of Laponite microstructure in the bulk of the
suspension. This microstructural change correlates with a decrease in
viscosity and therefore a decrease in $\Delta P_{s}/L$ with increase
in $Q$.  Such a behaviour, akin to a shear thinning fluid, will always
yield an increase in the steady (or saturated) state pressure drop
with increase in the flow rate in contrast to the observed behavior
over here. Further, we anticipate that such flow-induced
microstructural breakage will occur locally - thus, an increase in
tube length should result in greater microstructural change and lower
$\Delta P_{s}/L$. In this situation, $\Delta P_{s}/L$ will not be
independent of the tube length, which is inconsistent with the
behavior observed in fig.~\ref{delp-lq}b. Therefore, the decrease in
$\Delta P_{s}/L$ with $Q$ is not a consequence of flow-induced changes
in Laponite microstructure in the bulk of the tube. We now consider
the possibility that slip of Laponite suspensions at the tube walls
reduces the pressure drop. Similar behavior, i.e. increase in the flow
rate under constant pressure gradient, been observed previously for
pressure-driven flow of Newtonian liquids~\cite{cheng02} and
visco-plastic fluids~\cite{ortega16}.

\begin{figure} \includegraphics[scale=0.6]{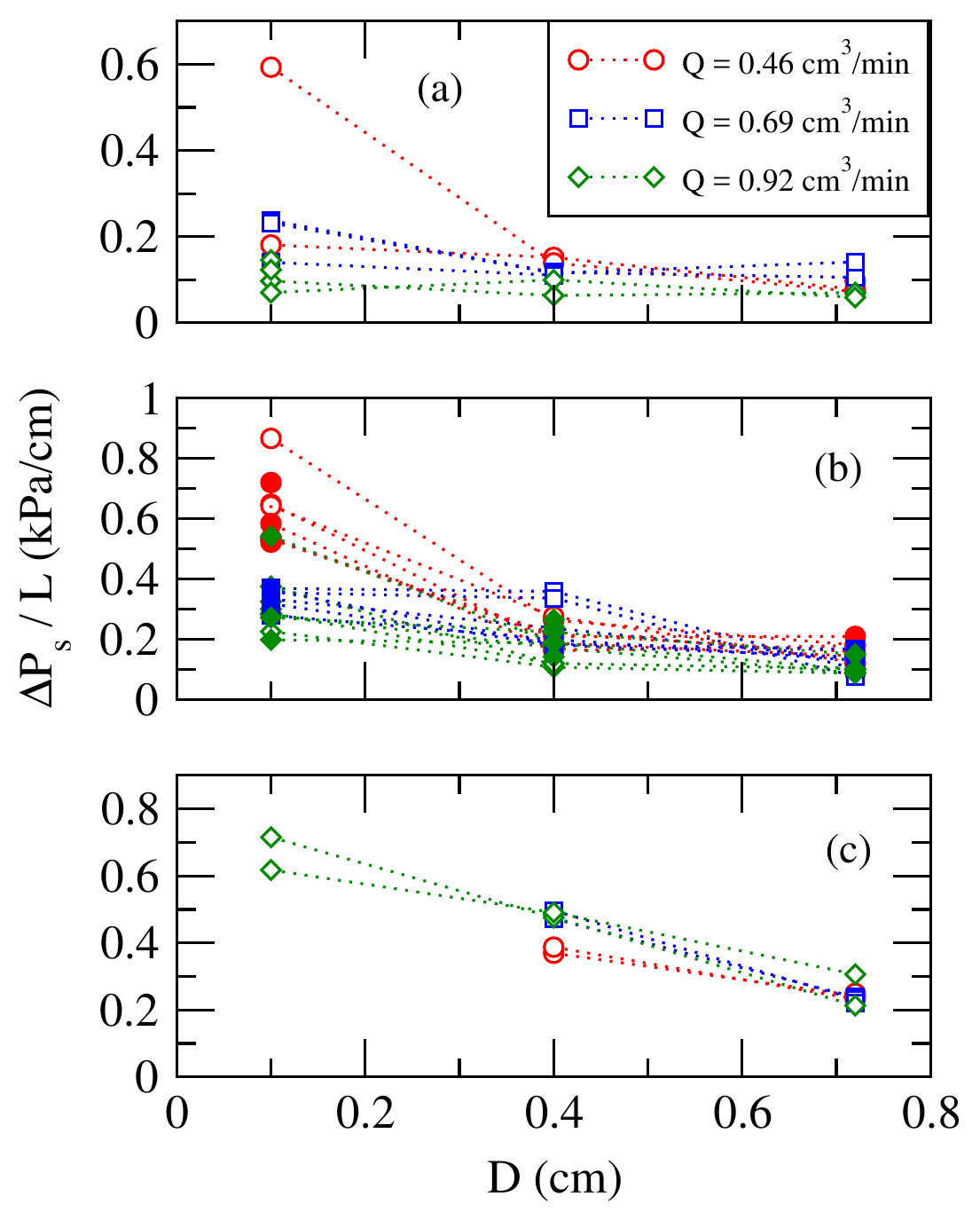}
  \caption{(a) Variation of saturated pressure drop per unit length
    ($\Delta P_{s}/L$) with tube diameter ($D$) for three flow rates
    ($Q$).  Panels (a), (b) and (c) represent, respectively, data
    obtained for salt concentrations, $c_{s} = 6$ mM, $c_{s} = 9$ mM
    and $c_{s} = 12$ mM. Open circles represent data for $L = 23$ cm,
    filled circles represent data for $L = 53$ cm, open squares and
    diamonds represent data for $L = 15.5$ cms and filed squares and
    diamonds represent data for $L = 30$ cm.}
  \label{delp-d}
\end{figure}

\subsection{Effect of tube diameter and salt concentration on
  saturated pressure drop}

Figure~\ref{delp-d} shows the variation of $\Delta P_{s}/L$ with tube
diameter for three different flow rates and three different salt
concentrations. The values of $\Delta P_{s}/L$ decrease with increase
in tube diameter at all flow rates and salt concentrations
employed. This effect is more pronounced at smaller tube
diameters. The inverse dependence is in line with the behavior
expected during the Poiseuille flow, i.e. smaller the cross-sectional
area available for flow, higher is the pressure drop needed to achieve
the same flow rate. Secondly, the magnitude of saturated pressure drop
increases significantly with increase in salt concentration. Now,
increase in salt concentration significantly accelerates ageing of
Laponite suspensions, thereby increasing its effective
viscosity~\cite{ruzicka11}. Naturally, the pressure drop required to
pump the fluid at a constant rate is expected to increase with
increase in the effective viscosity, in line with that expected from
the Poiseuille flow. The effective viscosity rises rapidly with salt
concentration, so that pressure drop measurements could be obtained
only at the highest flow rate employed for experiments using the
smallest tube diameter ($D = 0.1$ cm) at the highest salt
concentration employed ($c_{s} =$ 12 mM).

The parametric dependence described above shows distinct dependence of
the measured pressure drop on three variables, namely flow rate, tube
diameter and salt concentration.  The salt content tends to influence
the viscosity while the flow cross-section is governed
by the tube diameter. However, both these effects seem to occur on the
backdrop of flow slippage as evidenced by the observed dependence on
employed flow rate. This encourages us to seek a non-linear
dependence, akin to a scaling law relating these three variables with
the saturated pressure drop. In the following we attempt to obtain a
scaling relation from the experimental observations.

\subsection{Scaling behavior}

The Laponite suspension exhibits significant thixotropy, i.e. its
state evolves continuously with waiting time. A steady state is, thus,
not achievable in such a system, thereby precluding the existence of a
stress constitutive equation comprising a steady shear
viscosity. However, the experimental observations exhibit steady (or
saturated) state pressure drop following initial transient period (see
fig.~\ref{s-pulse}).  Given our primary interest in understanding the
behavior of steady pressure drop, we assume a unidirectional, steady
state flow of Laponite suspension through the tube.  Typically, a
viscous laponite suspension exhibits a yield stress ($\tau_{y}$) and a
shear flow post yielding. To simplify the representation of the
observed behavior, we consider Bingham fluid like behavior for which
the shear stress is expressed as $\tau = \tau_{y} + \mu \dot{\gamma}$,
where $\dot{\gamma}$ is the shear rate, $\mu$ is the shear viscosity.
Using these assumptions and following the previous treatment for flow
of Newtonian liquids~\cite{lauga03} and visco-plastic
fluids~\cite{kalyon05} through a cylindrical tube, the volumetric flow
rate can be obtained as
\begin{equation}
  Q = \frac{\pi D^{4}}{64} \left (\frac{\Delta P_{s}}{\mu L}
  \right) \left (1-\lambda+\frac{2\delta}{D}\right)^{2}
  \left [1- 2 \left (\frac{(1-\lambda)^{2}}{4}+\frac{(1-\lambda)}{3}  \lambda \right) \right]
  \label{eqn:flow-rate}
\end{equation}
where $\lambda = \tau_{y}/(\Delta P_{s} D/4L)$ and $\delta$ is the
slip length. The overall form of the Eq.~\ref{eq:flow-rate} is
consistent with that obtained previously for a visco-plastic
fluid~\cite{kalyon05}.

\begin{figure} \includegraphics[scale=0.55]{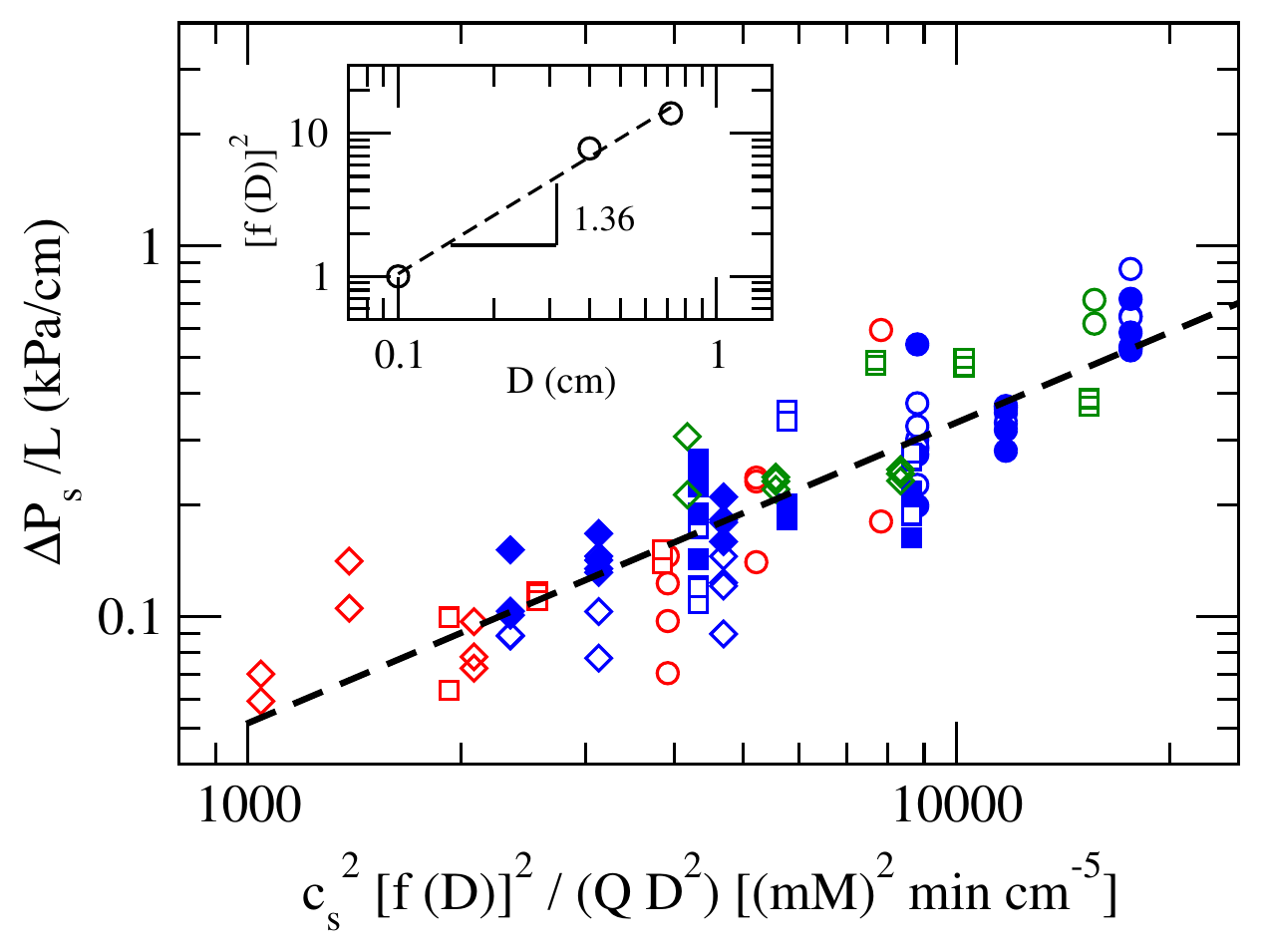}
  \caption{Scaling of saturated pressure drop per unit length
    ($\Delta P_{s}/L$) with different experimental parameters
    investigated. Inset: Variation of shift factor $[f(D)]^{2}$ with
    tube diameter.}
  \label{scaling}
\end{figure}

The value of $(\Delta P_{s} D/4L)$, i.e. the wall shear stress, as
measured from the experiments is of the order $[O(10 Pa)]$, similar to
the yield stress of Laponite dispersions of comparable concentration
and ionic strength~\cite{au15,lin19}. This implies that the material
does not shear anywhere in radial direction and moves as a plug,
viz. $\lambda \rightarrow 1$. This
is clearly observed through flow visualization (see section IV in
supplementary information and the accompanying movie file).  The material
yielding is then the primary reason for slip in present system which
is consistent with previous reports~\cite{poumaera14}.  Based on these
arguments, we consider slip as the predominant behavior which,
thereby, leads to simplification of eq.~(\ref{eq:flow-rate}) as
\begin{equation}
  Q = \frac{\pi}{16} \frac{\Delta P_{s}}{L} \frac{D^{2}
    \delta^{2}}{\mu}
  \label{eqn:dpexpr}
\end{equation}
In this equation, all quantities except $\delta$, are either known,
measured or can be estimated. We infer the behavior of $\delta$ based
on our experiments. Using a linear dependence i.e $\delta \propto Q$
in Eq.~\ref{eqn:dpexpr}, we observe that the experimentally obtained
inverse relation between $\Delta P_{s}$ and $Q$ is recovered. We
reiterate that the linear dependence of $\delta$ on $Q$ is invoked to
rationalize the data and is not based on any reasoning from the
literature. We do note, however, that experiments by Zhu and
Granick~\cite{zhu01} for Newtonian liquids do show that slip length
increases with shear rate. Further, neither can the
dependence of slip length on system size be obtained from
the literature~\cite{watanabe99,cheng02,li19}. We, thus,
consider $\delta \propto Q/f(D)$, where the exact functional
dependence on tube diameter is obtained from a fit to the experimental
data. Finally, the extensive literature on the effect of ionic
strength on Laponite aging indicates that the dependence on salt
concentration is highly nonlinear ~\cite{ruzicka11,joshi18a,joshi18b}.
Again out of convenience we invoke a non-linear relation where the
viscosity varies with the square of salt concentration, i.e.
$\mu \propto c_{s}^{2}$.  Using these assumptions,
Eq.~\ref{eqn:dpexpr} can be expressed as
\begin{equation}
  \frac{\Delta P_{s}}{L} \propto \frac{c_{s}^{2}}{Q D^{2}} [f(D)]^{2}
  \label{eqn:scal}
\end{equation}

The experimental results for all flow rates, tube diameters and salt
concentrations studied are shown in fig.~\ref{scaling} (main panel)
following Eq.~\ref{eqn:scal}. We adjust $f(D)$ so as to collapse the
data on a universal curve by shifting individual data sets.  We
observe that $[f(D)]^{2}$ shows a power-law dependence on tube
diameter ($D$) with an exponent $1.36$ as shown in
fig~\ref{scaling}(inset).  Collapse of data from experiments carried 
over a wide range of parameters, on a master curve is remarkable given the
complexity of microstructure formation in thixotropic Laponite
suspensions.  From a fit to the data, we obtain the relation
$(\Delta P_{s}/L) \propto c_{s}^{2} Q^{-1} D^{-0.64}$. Further, our
data suggests an empirical correlation for the slip length, given as
$\delta \propto Q/D^{0.68}$. Though not identical to our experimental
system, we note that the slip length has been theoretically shown to
scale as $D^{-1}$ for the flow of water through nanopores~\cite{li19}.

\section{Conclusions}

In summary, we have investigated the slip behavior for pressure driven
flow of a thixotropic material through a cylindrical tube.  While the
Laponite suspension is thixotropic, we obtain steady state flow
behavior over a range of experimental conditions. The steady state is
envisaged as a balance between the inherent structure formation of
Laponite and possible breakage due to flow. The steady state is
represented by near time independent (or saturated) pressure values
measured during the flow of suspension through the tube.

Remarkably, the saturated pressure drop shows an inverse relation with
the flow rate, in contrast to our expectation based on the Poiseuille
relation.  These observations can be rationalised by invoking slip of
Laponite suspensions at the tube walls. We show that the observed
experimental results can be accounted for if a linear dependence is
assumed between the slip length and flow rate and the material can be
described using Bingham constitutive equation. The observed scaling
behavior indicate that the slip length varies linearly with the flow
rate and inversely with tube diameter.

It is to be, however, noted that the scaling behavior is obtained
purely based on a fit to the data. The assumed variations of the key
variables cannot be obtained from the literature and their deduction
through independent measurements is not within the scope of this
work. The reasonably good scaling behavior suggests that these adhoc
assumptions are not without merit.  The remarkably simple dependence
of the flow behavior of a complex thixotropic material should pave the
way for a more involved theoretical treatment of flowing thixotropic
materials.

\section*{Supplementary Material}

Details about the design and characterisation of various flow accessories
used in experiments and flow visualisation methodology are provided as
supplementary material for ready reference.

\begin{acknowledgments}

  AVO gratefully acknowledges the financial support from Science \&
  Engineering Research Board, India (Grant No. SB/S3/CE/017/2015).

\end{acknowledgments}

\section*{Data Availability}

The data that support the findings of this study are available from
the corresponding author upon reasonable request.

\bibliography{ref-slip-flow}

\providecommand{\noopsort}[1]{}\providecommand{\singleletter}[1]{#1}%
\begin{thebibliography}{40}%
\makeatletter
\providecommand \@ifxundefined [1]{%
 \@ifx{#1\undefined}
}%
\providecommand \@ifnum [1]{%
 \ifnum #1\expandafter \@firstoftwo
 \else \expandafter \@secondoftwo
 \fi
}%
\providecommand \@ifx [1]{%
 \ifx #1\expandafter \@firstoftwo
 \else \expandafter \@secondoftwo
 \fi
}%
\providecommand \natexlab [1]{#1}%
\providecommand \enquote  [1]{``#1''}%
\providecommand \bibnamefont  [1]{#1}%
\providecommand \bibfnamefont [1]{#1}%
\providecommand \citenamefont [1]{#1}%
\providecommand \href@noop [0]{\@secondoftwo}%
\providecommand \href [0]{\begingroup \@sanitize@url \@href}%
\providecommand \@href[1]{\@@startlink{#1}\@@href}%
\providecommand \@@href[1]{\endgroup#1\@@endlink}%
\providecommand \@sanitize@url [0]{\catcode `\\12\catcode `\$12\catcode
  `\&12\catcode `\#12\catcode `\^12\catcode `\_12\catcode `\%12\relax}%
\providecommand \@@startlink[1]{}%
\providecommand \@@endlink[0]{}%
\providecommand \url  [0]{\begingroup\@sanitize@url \@url }%
\providecommand \@url [1]{\endgroup\@href {#1}{\urlprefix }}%
\providecommand \urlprefix  [0]{URL }%
\providecommand \Eprint [0]{\href }%
\providecommand \doibase [0]{http://dx.doi.org/}%
\providecommand \selectlanguage [0]{\@gobble}%
\providecommand \bibinfo  [0]{\@secondoftwo}%
\providecommand \bibfield  [0]{\@secondoftwo}%
\providecommand \translation [1]{[#1]}%
\providecommand \BibitemOpen [0]{}%
\providecommand \bibitemStop [0]{}%
\providecommand \bibitemNoStop [0]{.\EOS\space}%
\providecommand \EOS [0]{\spacefactor3000\relax}%
\providecommand \BibitemShut  [1]{\csname bibitem#1\endcsname}%
\let\auto@bib@innerbib\@empty
\bibitem [{\citenamefont {Schnell}(1956)}]{schnell56}%
  \BibitemOpen
  \bibfield  {author} {\bibinfo {author} {\bibfnamefont {E.}~\bibnamefont
  {Schnell}},\ }\bibfield  {title} {\enquote {\bibinfo {title} {Slippage of
  water over nonwettable surfaces},}\ }\href@noop {} {\bibfield  {journal}
  {\bibinfo  {journal} {J. Appl. Phys.}\ }\textbf {\bibinfo {volume} {27}},\
  \bibinfo {pages} {1149} (\bibinfo {year} {1956})}\BibitemShut {NoStop}%
\bibitem [{\citenamefont {Watanabe}\ and\ \citenamefont
  {nd~H.~Udagawa}(1999)}]{watanabe99}%
  \BibitemOpen
  \bibfield  {author} {\bibinfo {author} {\bibfnamefont {K.}~\bibnamefont
  {Watanabe}}\ and\ \bibinfo {author} {\bibfnamefont {Y.~U.}\ \bibnamefont
  {nd~H.~Udagawa}},\ }\bibfield  {title} {\enquote {\bibinfo {title} {Drag
  reduction of newtonian fluid in a circular pipe with a higly water-repellent
  wall},}\ }\href@noop {} {\bibfield  {journal} {\bibinfo  {journal} {J. Fluid
  Mech.}\ }\textbf {\bibinfo {volume} {381}},\ \bibinfo {pages} {225--238}
  (\bibinfo {year} {1999})}\BibitemShut {NoStop}%
\bibitem [{\citenamefont {Tretheway}\ and\ \citenamefont
  {Meinhart}(2002)}]{tretheway02}%
  \BibitemOpen
  \bibfield  {author} {\bibinfo {author} {\bibfnamefont {D.~C.}\ \bibnamefont
  {Tretheway}}\ and\ \bibinfo {author} {\bibfnamefont {C.~D.}\ \bibnamefont
  {Meinhart}},\ }\bibfield  {title} {\enquote {\bibinfo {title} {Apparent fluid
  slip at hydrophobic microchannel walls},}\ }\href@noop {} {\bibfield
  {journal} {\bibinfo  {journal} {Phys. Fluids}\ }\textbf {\bibinfo {volume}
  {14}},\ \bibinfo {pages} {L9} (\bibinfo {year} {2002})}\BibitemShut {NoStop}%
\bibitem [{\citenamefont {Li}\ \emph {et~al.}(2019)\citenamefont {Li},
  \citenamefont {Su}, \citenamefont {Wang}, \citenamefont {Sheng},\ and\
  \citenamefont {Wang}}]{li19}%
  \BibitemOpen
  \bibfield  {author} {\bibinfo {author} {\bibfnamefont {L.}~\bibnamefont
  {Li}}, \bibinfo {author} {\bibfnamefont {Y.}~\bibnamefont {Su}}, \bibinfo
  {author} {\bibfnamefont {H.}~\bibnamefont {Wang}}, \bibinfo {author}
  {\bibfnamefont {G.}~\bibnamefont {Sheng}}, \ and\ \bibinfo {author}
  {\bibfnamefont {W.}~\bibnamefont {Wang}},\ }\bibfield  {title} {\enquote
  {\bibinfo {title} {A new slip length model for enhanced water flow coupling
  molecular interaction, pore dimension, wall roughness and temperature},}\
  }\href@noop {} {\bibfield  {journal} {\bibinfo  {journal} {Adv. Polymer
  Technol.}\ }\textbf {\bibinfo {volume} {2019}},\ \bibinfo {pages} {6424012}
  (\bibinfo {year} {2019})}\BibitemShut {NoStop}%
\bibitem [{\citenamefont {Churaev}, \citenamefont {Sobolev},\ and\
  \citenamefont {Somov}(1984)}]{churaev84}%
  \BibitemOpen
  \bibfield  {author} {\bibinfo {author} {\bibfnamefont {N.~V.}\ \bibnamefont
  {Churaev}}, \bibinfo {author} {\bibfnamefont {V.~D.}\ \bibnamefont
  {Sobolev}}, \ and\ \bibinfo {author} {\bibfnamefont {A.~N.}\ \bibnamefont
  {Somov}},\ }\bibfield  {title} {\enquote {\bibinfo {title} {Slippage of
  liquids over lyophobic solid surfaces},}\ }\href@noop {} {\bibfield
  {journal} {\bibinfo  {journal} {J. Coll. Int. Sci.}\ }\textbf {\bibinfo
  {volume} {97}},\ \bibinfo {pages} {574} (\bibinfo {year} {1984})}\BibitemShut
  {NoStop}%
\bibitem [{\citenamefont {Aghdam}\ and\ \citenamefont
  {Ricco}(2016)}]{aghdam16}%
  \BibitemOpen
  \bibfield  {author} {\bibinfo {author} {\bibfnamefont {S.~K.}\ \bibnamefont
  {Aghdam}}\ and\ \bibinfo {author} {\bibfnamefont {P.}~\bibnamefont {Ricco}},\
  }\bibfield  {title} {\enquote {\bibinfo {title} {Laminar and turbulent flows
  over hydrophobic surfaces with shear-dependent slip length},}\ }\href@noop {}
  {\bibfield  {journal} {\bibinfo  {journal} {Phys. Fluids}\ }\textbf {\bibinfo
  {volume} {28}},\ \bibinfo {pages} {035109} (\bibinfo {year}
  {2016})}\BibitemShut {NoStop}%
\bibitem [{\citenamefont {Cheng}\ and\ \citenamefont
  {Giordano}(2002)}]{cheng02}%
  \BibitemOpen
  \bibfield  {author} {\bibinfo {author} {\bibfnamefont {J.~T.}\ \bibnamefont
  {Cheng}}\ and\ \bibinfo {author} {\bibfnamefont {N.}~\bibnamefont
  {Giordano}},\ }\bibfield  {title} {\enquote {\bibinfo {title} {Fluid flow
  through nanometer-scale channels},}\ }\href@noop {} {\bibfield  {journal}
  {\bibinfo  {journal} {Phys. Rev. E}\ }\textbf {\bibinfo {volume} {65}},\
  \bibinfo {pages} {031206} (\bibinfo {year} {2002})}\BibitemShut {NoStop}%
\bibitem [{\citenamefont {Lauga}\ and\ \citenamefont {Stone}(2003)}]{lauga03}%
  \BibitemOpen
  \bibfield  {author} {\bibinfo {author} {\bibfnamefont {E.}~\bibnamefont
  {Lauga}}\ and\ \bibinfo {author} {\bibfnamefont {H.~A.}\ \bibnamefont
  {Stone}},\ }\bibfield  {title} {\enquote {\bibinfo {title} {Effective slip in
  pressure-driven stokes flow},}\ }\href@noop {} {\bibfield  {journal}
  {\bibinfo  {journal} {J. Fluid Mech.}\ }\textbf {\bibinfo {volume} {489}},\
  \bibinfo {pages} {55--77} (\bibinfo {year} {2003})}\BibitemShut {NoStop}%
\bibitem [{\citenamefont {Kalyon}(2005)}]{kalyon05}%
  \BibitemOpen
  \bibfield  {author} {\bibinfo {author} {\bibfnamefont {D.~K.}\ \bibnamefont
  {Kalyon}},\ }\bibfield  {title} {\enquote {\bibinfo {title} {Apparent slip
  and viscoplasticity of concentrated suspensions},}\ }\href@noop {} {\bibfield
   {journal} {\bibinfo  {journal} {J. Rheol.}\ }\textbf {\bibinfo {volume}
  {49}},\ \bibinfo {pages} {621} (\bibinfo {year} {2005})}\BibitemShut
  {NoStop}%
\bibitem [{\citenamefont {Kalyon}\ and\ \citenamefont
  {Malik}(2012)}]{kalyon12}%
  \BibitemOpen
  \bibfield  {author} {\bibinfo {author} {\bibfnamefont {D.~M.}\ \bibnamefont
  {Kalyon}}\ and\ \bibinfo {author} {\bibfnamefont {M.}~\bibnamefont {Malik}},\
  }\bibfield  {title} {\enquote {\bibinfo {title} {Axial laminar flow of
  viscoplastic fluids in a concentric annulus subject to wall slip},}\
  }\href@noop {} {\bibfield  {journal} {\bibinfo  {journal} {Rheol. Acta}\
  }\textbf {\bibinfo {volume} {51}},\ \bibinfo {pages} {805} (\bibinfo {year}
  {2012})}\BibitemShut {NoStop}%
\bibitem [{\citenamefont {Ortega-Avila}\ \emph {et~al.}(2016)\citenamefont
  {Ortega-Avila}, \citenamefont {P\'{e}rez-Gonz\'{a}lez}, \citenamefont
  {Mar\'{i}n-Santib\'{a}\'{n}ez}, \citenamefont {Rodr\'{i}guez-Gonz\'{a}lez},
  \citenamefont {Aktas}, \citenamefont {Malik},\ and\ \citenamefont
  {Kalyon}}]{ortega16}%
  \BibitemOpen
  \bibfield  {author} {\bibinfo {author} {\bibfnamefont {J.~F.}\ \bibnamefont
  {Ortega-Avila}}, \bibinfo {author} {\bibfnamefont {J.}~\bibnamefont
  {P\'{e}rez-Gonz\'{a}lez}}, \bibinfo {author} {\bibfnamefont {B.~M.}\
  \bibnamefont {Mar\'{i}n-Santib\'{a}\'{n}ez}}, \bibinfo {author}
  {\bibfnamefont {F.}~\bibnamefont {Rodr\'{i}guez-Gonz\'{a}lez}}, \bibinfo
  {author} {\bibfnamefont {S.}~\bibnamefont {Aktas}}, \bibinfo {author}
  {\bibfnamefont {M.}~\bibnamefont {Malik}}, \ and\ \bibinfo {author}
  {\bibfnamefont {D.~M.}\ \bibnamefont {Kalyon}},\ }\bibfield  {title}
  {\enquote {\bibinfo {title} {Axial annular flow of a viscoplastic microgel
  with wall slip},}\ }\href@noop {} {\bibfield  {journal} {\bibinfo  {journal}
  {J. Rheol.}\ }\textbf {\bibinfo {volume} {60}},\ \bibinfo {pages} {503}
  (\bibinfo {year} {2016})}\BibitemShut {NoStop}%
\bibitem [{\citenamefont {Poumaere}\ \emph {et~al.}(2014)\citenamefont
  {Poumaere}, \citenamefont {Moyers-Gonz\'{a}lez}, \citenamefont {Castelain},\
  and\ \citenamefont {Burghelea}}]{poumaera14}%
  \BibitemOpen
  \bibfield  {author} {\bibinfo {author} {\bibfnamefont {A.}~\bibnamefont
  {Poumaere}}, \bibinfo {author} {\bibfnamefont {M.}~\bibnamefont
  {Moyers-Gonz\'{a}lez}}, \bibinfo {author} {\bibfnamefont {C.}~\bibnamefont
  {Castelain}}, \ and\ \bibinfo {author} {\bibfnamefont {T.}~\bibnamefont
  {Burghelea}},\ }\bibfield  {title} {\enquote {\bibinfo {title} {Unsteady
  laminar flows of a carbopol gel in the presence of wall slip},}\ }\href@noop
  {} {\bibfield  {journal} {\bibinfo  {journal} {J. non-Newt. Fluid Mech.}\
  }\textbf {\bibinfo {volume} {205}},\ \bibinfo {pages} {28} (\bibinfo {year}
  {2014})}\BibitemShut {NoStop}%
\bibitem [{\citenamefont {Haase}\ \emph {et~al.}(2017)\citenamefont {Haase},
  \citenamefont {Wood}, \citenamefont {Sprakel},\ and\ \citenamefont
  {Lammertink}}]{haase17}%
  \BibitemOpen
  \bibfield  {author} {\bibinfo {author} {\bibfnamefont {A.~S.}\ \bibnamefont
  {Haase}}, \bibinfo {author} {\bibfnamefont {J.~A.}\ \bibnamefont {Wood}},
  \bibinfo {author} {\bibfnamefont {L.~M.~J.}\ \bibnamefont {Sprakel}}, \ and\
  \bibinfo {author} {\bibfnamefont {R.~G.~H.}\ \bibnamefont {Lammertink}},\
  }\bibfield  {title} {\enquote {\bibinfo {title} {Inelastic non-newtonian flow
  over heterogeneously slippery surfaces},}\ }\href@noop {} {\bibfield
  {journal} {\bibinfo  {journal} {Phys. Rev. E}\ }\textbf {\bibinfo {volume}
  {95}},\ \bibinfo {pages} {023105} (\bibinfo {year} {2017})}\BibitemShut
  {NoStop}%
\bibitem [{\citenamefont {Aiyejina}\ \emph {et~al.}(2011)\citenamefont
  {Aiyejina}, \citenamefont {Chakrabarti}, \citenamefont {Pilgrim},\ and\
  \citenamefont {Sastry}}]{sastry11}%
  \BibitemOpen
  \bibfield  {author} {\bibinfo {author} {\bibfnamefont {A.}~\bibnamefont
  {Aiyejina}}, \bibinfo {author} {\bibfnamefont {D.~P.}\ \bibnamefont
  {Chakrabarti}}, \bibinfo {author} {\bibfnamefont {A.}~\bibnamefont
  {Pilgrim}}, \ and\ \bibinfo {author} {\bibfnamefont {M.~K.~S.}\ \bibnamefont
  {Sastry}},\ }\bibfield  {title} {\enquote {\bibinfo {title} {Wax formation in
  oil pipelines: A critical review},}\ }\href@noop {} {\bibfield  {journal}
  {\bibinfo  {journal} {Int. J. Multi, Flow}\ }\textbf {\bibinfo {volume}
  {37}},\ \bibinfo {pages} {671--694} (\bibinfo {year} {2011})}\BibitemShut
  {NoStop}%
\bibitem [{\citenamefont {van Beerkel}\ \emph {et~al.}(2005)\citenamefont {van
  Beerkel}, \citenamefont {van Marle}, \citenamefont {Groen},\ and\
  \citenamefont {Bruno}}]{berkel05}%
  \BibitemOpen
  \bibfield  {author} {\bibinfo {author} {\bibfnamefont {A.~M.}\ \bibnamefont
  {van Beerkel}}, \bibinfo {author} {\bibfnamefont {J.}~\bibnamefont {van
  Marle}}, \bibinfo {author} {\bibfnamefont {A.~K.}\ \bibnamefont {Groen}}, \
  and\ \bibinfo {author} {\bibfnamefont {M.~J.}\ \bibnamefont {Bruno}},\
  }\bibfield  {title} {\enquote {\bibinfo {title} {Mechanisms of biliary stent
  clogging},}\ }\href@noop {} {\bibfield  {journal} {\bibinfo  {journal}
  {Endoscopy}\ }\textbf {\bibinfo {volume} {37}},\ \bibinfo {pages} {729--734}
  (\bibinfo {year} {2005})}\BibitemShut {NoStop}%
\bibitem [{\citenamefont {Donelli}\ \emph {et~al.}(2007)\citenamefont
  {Donelli}, \citenamefont {Guaglianone}, \citenamefont {Rosa}, \citenamefont
  {Fiocca},\ and\ \citenamefont {Basoli}}]{donelli07}%
  \BibitemOpen
  \bibfield  {author} {\bibinfo {author} {\bibfnamefont {G.}~\bibnamefont
  {Donelli}}, \bibinfo {author} {\bibfnamefont {E.}~\bibnamefont
  {Guaglianone}}, \bibinfo {author} {\bibfnamefont {R.~D.}\ \bibnamefont
  {Rosa}}, \bibinfo {author} {\bibfnamefont {F.}~\bibnamefont {Fiocca}}, \ and\
  \bibinfo {author} {\bibfnamefont {A.}~\bibnamefont {Basoli}},\ }\bibfield
  {title} {\enquote {\bibinfo {title} {Plastic biliary stent occlusion: Factors
  involved and possible preventiva approaches},}\ }\href@noop {} {\bibfield
  {journal} {\bibinfo  {journal} {Clinical Med. Res.}\ }\textbf {\bibinfo
  {volume} {5}},\ \bibinfo {pages} {53--60} (\bibinfo {year}
  {2007})}\BibitemShut {NoStop}%
\bibitem [{\citenamefont {Jamali}, \citenamefont {McKinley},\ and\
  \citenamefont {Armstrong}(2017)}]{jamali17}%
  \BibitemOpen
  \bibfield  {author} {\bibinfo {author} {\bibfnamefont {S.}~\bibnamefont
  {Jamali}}, \bibinfo {author} {\bibfnamefont {G.~H.}\ \bibnamefont
  {McKinley}}, \ and\ \bibinfo {author} {\bibfnamefont {R.~C.}\ \bibnamefont
  {Armstrong}},\ }\bibfield  {title} {\enquote {\bibinfo {title}
  {Microstructural rearrangements and their rheological implications in a model
  thixotropic elastoviscoplastic fluid},}\ }\href@noop {} {\bibfield  {journal}
  {\bibinfo  {journal} {Phys. Rev. Lett.}\ }\textbf {\bibinfo {volume} {118}},\
  \bibinfo {pages} {048003} (\bibinfo {year} {2017})}\BibitemShut {NoStop}%
\bibitem [{\citenamefont {Cunha}, \citenamefont {de~Souza~Mendes},\ and\
  \citenamefont {Siqueira}(2020)}]{cunha20}%
  \BibitemOpen
  \bibfield  {author} {\bibinfo {author} {\bibfnamefont {J.~P.}\ \bibnamefont
  {Cunha}}, \bibinfo {author} {\bibfnamefont {P.~R.}\ \bibnamefont
  {de~Souza~Mendes}}, \ and\ \bibinfo {author} {\bibfnamefont {I.~R.}\
  \bibnamefont {Siqueira}},\ }\bibfield  {title} {\enquote {\bibinfo {title}
  {Pressure-driven flows of a thixotropic viscoplastic material: Performance of
  a novel fluidity-based constitutive model},}\ }\href@noop {} {\bibfield
  {journal} {\bibinfo  {journal} {Phys. Fluids}\ }\textbf {\bibinfo {volume}
  {32}},\ \bibinfo {pages} {123104} (\bibinfo {year} {2020})}\BibitemShut
  {NoStop}%
\bibitem [{\citenamefont {Bonn}\ \emph {et~al.}(1999)\citenamefont {Bonn},
  \citenamefont {Kellay}, \citenamefont {Tanaka}, \citenamefont {Wegdam},\ and\
  \citenamefont {Meunier}}]{bonn99}%
  \BibitemOpen
  \bibfield  {author} {\bibinfo {author} {\bibfnamefont {D.}~\bibnamefont
  {Bonn}}, \bibinfo {author} {\bibfnamefont {H.}~\bibnamefont {Kellay}},
  \bibinfo {author} {\bibfnamefont {H.}~\bibnamefont {Tanaka}}, \bibinfo
  {author} {\bibfnamefont {G.}~\bibnamefont {Wegdam}}, \ and\ \bibinfo {author}
  {\bibfnamefont {J.}~\bibnamefont {Meunier}},\ }\bibfield  {title} {\enquote
  {\bibinfo {title} {Laponite: What is the difference between a glass and a
  gel?}}\ }\href@noop {} {\bibfield  {journal} {\bibinfo  {journal} {Langmuir}\
  }\textbf {\bibinfo {volume} {15}},\ \bibinfo {pages} {7534--7536} (\bibinfo
  {year} {1999})}\BibitemShut {NoStop}%
\bibitem [{\citenamefont {Knaebel}\ \emph {et~al.}(2000)\citenamefont
  {Knaebel}, \citenamefont {Bellour}, \citenamefont {Munch}, \citenamefont
  {Viasnoff}, \citenamefont {Lequeux},\ and\ \citenamefont
  {Harden}}]{knaebel00}%
  \BibitemOpen
  \bibfield  {author} {\bibinfo {author} {\bibfnamefont {A.}~\bibnamefont
  {Knaebel}}, \bibinfo {author} {\bibfnamefont {M.}~\bibnamefont {Bellour}},
  \bibinfo {author} {\bibfnamefont {J.~P.}\ \bibnamefont {Munch}}, \bibinfo
  {author} {\bibfnamefont {V.}~\bibnamefont {Viasnoff}}, \bibinfo {author}
  {\bibfnamefont {F.}~\bibnamefont {Lequeux}}, \ and\ \bibinfo {author}
  {\bibfnamefont {J.~L.}\ \bibnamefont {Harden}},\ }\bibfield  {title}
  {\enquote {\bibinfo {title} {Aging behavior of laponite clay particle
  suspensions},}\ }\href@noop {} {\bibfield  {journal} {\bibinfo  {journal}
  {Eur. Phys. Lett.}\ }\textbf {\bibinfo {volume} {52}},\ \bibinfo {pages} {73}
  (\bibinfo {year} {2000})}\BibitemShut {NoStop}%
\bibitem [{\citenamefont {Bonn}\ \emph {et~al.}(2002)\citenamefont {Bonn},
  \citenamefont {Tanase}, \citenamefont {Abou}, \citenamefont {Tanaka},\ and\
  \citenamefont {Meunier}}]{bonn02}%
  \BibitemOpen
  \bibfield  {author} {\bibinfo {author} {\bibfnamefont {D.}~\bibnamefont
  {Bonn}}, \bibinfo {author} {\bibfnamefont {S.}~\bibnamefont {Tanase}},
  \bibinfo {author} {\bibfnamefont {B.}~\bibnamefont {Abou}}, \bibinfo {author}
  {\bibfnamefont {H.}~\bibnamefont {Tanaka}}, \ and\ \bibinfo {author}
  {\bibfnamefont {J.}~\bibnamefont {Meunier}},\ }\bibfield  {title} {\enquote
  {\bibinfo {title} {Laponite: Aging and shear rejuvenation of a colloidal
  glass},}\ }\href@noop {} {\bibfield  {journal} {\bibinfo  {journal} {Phys.
  Rev. Lett.}\ }\textbf {\bibinfo {volume} {89}},\ \bibinfo {pages} {015701}
  (\bibinfo {year} {2002})}\BibitemShut {NoStop}%
\bibitem [{\citenamefont {Ruzicka}\ \emph {et~al.}(2010)\citenamefont
  {Ruzicka}, \citenamefont {Zulian}, \citenamefont {Zaccarelli}, \citenamefont
  {Angelini}, \citenamefont {Sztucki}, \citenamefont {Moussaid},\ and\
  \citenamefont {Ruocco}}]{ruzicka10}%
  \BibitemOpen
  \bibfield  {author} {\bibinfo {author} {\bibfnamefont {B.}~\bibnamefont
  {Ruzicka}}, \bibinfo {author} {\bibfnamefont {L.}~\bibnamefont {Zulian}},
  \bibinfo {author} {\bibfnamefont {E.}~\bibnamefont {Zaccarelli}}, \bibinfo
  {author} {\bibfnamefont {R.}~\bibnamefont {Angelini}}, \bibinfo {author}
  {\bibfnamefont {M.}~\bibnamefont {Sztucki}}, \bibinfo {author} {\bibfnamefont
  {A.}~\bibnamefont {Moussaid}}, \ and\ \bibinfo {author} {\bibfnamefont
  {G.}~\bibnamefont {Ruocco}},\ }\bibfield  {title} {\enquote {\bibinfo {title}
  {Competing interactions in arrested states of colloidal clays},}\ }\href@noop
  {} {\bibfield  {journal} {\bibinfo  {journal} {Phys. Rev. Lett.}\ }\textbf
  {\bibinfo {volume} {104}},\ \bibinfo {pages} {085701} (\bibinfo {year}
  {2010})}\BibitemShut {NoStop}%
\bibitem [{\citenamefont {Ruzicka}\ and\ \citenamefont
  {Zaccarelli}(2011)}]{ruzicka11}%
  \BibitemOpen
  \bibfield  {author} {\bibinfo {author} {\bibfnamefont {B.}~\bibnamefont
  {Ruzicka}}\ and\ \bibinfo {author} {\bibfnamefont {E.}~\bibnamefont
  {Zaccarelli}},\ }\bibfield  {title} {\enquote {\bibinfo {title} {A fresh look
  at the laponite phase diagram},}\ }\href@noop {} {\bibfield  {journal}
  {\bibinfo  {journal} {Soft Mat.}\ }\textbf {\bibinfo {volume} {7}},\ \bibinfo
  {pages} {1268--1286} (\bibinfo {year} {2011})}\BibitemShut {NoStop}%
\bibitem [{\citenamefont {Jatav}\ and\ \citenamefont {Joshi}(2014)}]{joshi14}%
  \BibitemOpen
  \bibfield  {author} {\bibinfo {author} {\bibfnamefont {S.}~\bibnamefont
  {Jatav}}\ and\ \bibinfo {author} {\bibfnamefont {Y.~M.}\ \bibnamefont
  {Joshi}},\ }\bibfield  {title} {\enquote {\bibinfo {title} {Rheological
  signatures of gelation and effect of shear melting on aging colloidal
  suspension},}\ }\href@noop {} {\bibfield  {journal} {\bibinfo  {journal} {J.
  Rheol.}\ }\textbf {\bibinfo {volume} {58}},\ \bibinfo {pages} {1535--1554}
  (\bibinfo {year} {2014})}\BibitemShut {NoStop}%
\bibitem [{\citenamefont {Suman}\ and\ \citenamefont {Joshi}(2018)}]{joshi18a}%
  \BibitemOpen
  \bibfield  {author} {\bibinfo {author} {\bibfnamefont {K.}~\bibnamefont
  {Suman}}\ and\ \bibinfo {author} {\bibfnamefont {Y.~M.}\ \bibnamefont
  {Joshi}},\ }\bibfield  {title} {\enquote {\bibinfo {title} {Microstructure
  and soft glassy dynamics of aqueous laponite dispersion},}\ }\href@noop {}
  {\bibfield  {journal} {\bibinfo  {journal} {Langmuir}\ }\textbf {\bibinfo
  {volume} {34}},\ \bibinfo {pages} {13079--13103} (\bibinfo {year}
  {2018})}\BibitemShut {NoStop}%
\bibitem [{\citenamefont {Levitz}\ \emph {et~al.}(2000)\citenamefont {Levitz},
  \citenamefont {Lecolier}, \citenamefont {Mourchid}, \citenamefont
  {Delville},\ and\ \citenamefont {Lyonnard}}]{Levitz00}%
  \BibitemOpen
  \bibfield  {author} {\bibinfo {author} {\bibfnamefont {P.}~\bibnamefont
  {Levitz}}, \bibinfo {author} {\bibfnamefont {E.}~\bibnamefont {Lecolier}},
  \bibinfo {author} {\bibfnamefont {A.}~\bibnamefont {Mourchid}}, \bibinfo
  {author} {\bibfnamefont {A.}~\bibnamefont {Delville}}, \ and\ \bibinfo
  {author} {\bibfnamefont {S.}~\bibnamefont {Lyonnard}},\ }\bibfield  {title}
  {\enquote {\bibinfo {title} {Liquid-solid transition of laponite suspensions
  at very low ionic strength: Long-range electrostatic stabilisation of
  anisotropic colloids},}\ }\href@noop {} {\bibfield  {journal} {\bibinfo
  {journal} {Eur. Phys. Lett.}\ }\textbf {\bibinfo {volume} {49}},\ \bibinfo
  {pages} {672} (\bibinfo {year} {2000})}\BibitemShut {NoStop}%
\bibitem [{\citenamefont {Bonn}\ \emph {et~al.}(1998)\citenamefont {Bonn},
  \citenamefont {Tanaka}, \citenamefont {Wegdam}, \citenamefont {Kellay},\ and\
  \citenamefont {Meunier}}]{Bonn98}%
  \BibitemOpen
  \bibfield  {author} {\bibinfo {author} {\bibfnamefont {D.}~\bibnamefont
  {Bonn}}, \bibinfo {author} {\bibfnamefont {H.}~\bibnamefont {Tanaka}},
  \bibinfo {author} {\bibfnamefont {G.}~\bibnamefont {Wegdam}}, \bibinfo
  {author} {\bibfnamefont {H.}~\bibnamefont {Kellay}}, \ and\ \bibinfo {author}
  {\bibfnamefont {J.}~\bibnamefont {Meunier}},\ }\bibfield  {title} {\enquote
  {\bibinfo {title} {Aging of a colloidal “wigner” glass},}\ }\href@noop {}
  {\bibfield  {journal} {\bibinfo  {journal} {Eur. Phys. Lett.}\ }\textbf
  {\bibinfo {volume} {45}},\ \bibinfo {pages} {52} (\bibinfo {year}
  {1998})}\BibitemShut {NoStop}%
\bibitem [{\citenamefont {Tanaka}, \citenamefont {Meunier},\ and\ \citenamefont
  {Bonn}(2004)}]{Tanaka04}%
  \BibitemOpen
  \bibfield  {author} {\bibinfo {author} {\bibfnamefont {H.}~\bibnamefont
  {Tanaka}}, \bibinfo {author} {\bibfnamefont {J.}~\bibnamefont {Meunier}}, \
  and\ \bibinfo {author} {\bibfnamefont {D.}~\bibnamefont {Bonn}},\ }\bibfield
  {title} {\enquote {\bibinfo {title} {Nonergodic states of charged colloidal
  suspensions: Repulsive and attractive glasses and gels},}\ }\href@noop {}
  {\bibfield  {journal} {\bibinfo  {journal} {Phys. Rev. E.}\ }\textbf
  {\bibinfo {volume} {69}},\ \bibinfo {pages} {031404} (\bibinfo {year}
  {2004})}\BibitemShut {NoStop}%
\bibitem [{\citenamefont {Mourchid}\ \emph {et~al.}(1995)\citenamefont
  {Mourchid}, \citenamefont {Delville}, \citenamefont {Lambard}, \citenamefont
  {Lecolier},\ and\ \citenamefont {Levitz}}]{Mourchid95}%
  \BibitemOpen
  \bibfield  {author} {\bibinfo {author} {\bibfnamefont {A.}~\bibnamefont
  {Mourchid}}, \bibinfo {author} {\bibfnamefont {A.}~\bibnamefont {Delville}},
  \bibinfo {author} {\bibfnamefont {J.}~\bibnamefont {Lambard}}, \bibinfo
  {author} {\bibfnamefont {E.}~\bibnamefont {Lecolier}}, \ and\ \bibinfo
  {author} {\bibfnamefont {P.}~\bibnamefont {Levitz}},\ }\bibfield  {title}
  {\enquote {\bibinfo {title} {Phase diagram of colloidal dispersions of
  anisotropic charged particles: equilibrium properties, structure, and
  rheology of laponite suspensions},}\ }\href@noop {} {\bibfield  {journal}
  {\bibinfo  {journal} {Langmuir}\ }\textbf {\bibinfo {volume} {11}},\ \bibinfo
  {pages} {1942--1950} (\bibinfo {year} {1995})}\BibitemShut {NoStop}%
\bibitem [{\citenamefont {Pignon}, \citenamefont {Piau},\ and\ \citenamefont
  {Magnin}(1996)}]{Pignon96}%
  \BibitemOpen
  \bibfield  {author} {\bibinfo {author} {\bibfnamefont {F.}~\bibnamefont
  {Pignon}}, \bibinfo {author} {\bibfnamefont {J.~M.}\ \bibnamefont {Piau}}, \
  and\ \bibinfo {author} {\bibfnamefont {A.}~\bibnamefont {Magnin}},\
  }\bibfield  {title} {\enquote {\bibinfo {title} {Structure and pertinent
  length scale of a discotic clay gel},}\ }\href@noop {} {\bibfield  {journal}
  {\bibinfo  {journal} {Phys. Rev. Lett.}\ }\textbf {\bibinfo {volume} {76}},\
  \bibinfo {pages} {4857} (\bibinfo {year} {1996})}\BibitemShut {NoStop}%
\bibitem [{\citenamefont {Pignon}, \citenamefont {Magnin},\ and\ \citenamefont
  {Piau}(1997)}]{Pignon97}%
  \BibitemOpen
  \bibfield  {author} {\bibinfo {author} {\bibfnamefont {F.}~\bibnamefont
  {Pignon}}, \bibinfo {author} {\bibfnamefont {A.}~\bibnamefont {Magnin}}, \
  and\ \bibinfo {author} {\bibfnamefont {J.~M.}\ \bibnamefont {Piau}},\
  }\bibfield  {title} {\enquote {\bibinfo {title} {Butterfly light scattering
  pattern and rheology of a sheared thixotropic clay gel},}\ }\href@noop {}
  {\bibfield  {journal} {\bibinfo  {journal} {Phys. Rev. Lett.}\ }\textbf
  {\bibinfo {volume} {79}},\ \bibinfo {pages} {4689} (\bibinfo {year}
  {1997})}\BibitemShut {NoStop}%
\bibitem [{\citenamefont {Pignon}\ \emph {et~al.}(1997)\citenamefont {Pignon},
  \citenamefont {Magnin}, \citenamefont {Piau}, \citenamefont {B.~Cabane},\
  and\ \citenamefont {Diat}}]{PignonE97}%
  \BibitemOpen
  \bibfield  {author} {\bibinfo {author} {\bibfnamefont {F.}~\bibnamefont
  {Pignon}}, \bibinfo {author} {\bibfnamefont {A.}~\bibnamefont {Magnin}},
  \bibinfo {author} {\bibfnamefont {J.~M.}\ \bibnamefont {Piau}}, \bibinfo
  {author} {\bibfnamefont {P.~L.}\ \bibnamefont {B.~Cabane}}, \ and\ \bibinfo
  {author} {\bibfnamefont {O.}~\bibnamefont {Diat}},\ }\bibfield  {title}
  {\enquote {\bibinfo {title} {Yield stress thixotropic clay suspension:
  Investigations of structure by light, neutron, and x-ray scattering},}\
  }\href@noop {} {\bibfield  {journal} {\bibinfo  {journal} {Phys. Rev. E.}\
  }\textbf {\bibinfo {volume} {56}},\ \bibinfo {pages} {3281} (\bibinfo {year}
  {1997})}\BibitemShut {NoStop}%
\bibitem [{\citenamefont {Kroon}, \citenamefont {Wegdam},\ and\ \citenamefont
  {Sprik}(1996)}]{Kroon96}%
  \BibitemOpen
  \bibfield  {author} {\bibinfo {author} {\bibfnamefont {M.}~\bibnamefont
  {Kroon}}, \bibinfo {author} {\bibfnamefont {G.}~\bibnamefont {Wegdam}}, \
  and\ \bibinfo {author} {\bibfnamefont {R.}~\bibnamefont {Sprik}},\ }\bibfield
   {title} {\enquote {\bibinfo {title} {Dynamic light scattering studies on the
  sol-gel transition of a suspension of anisotropic colloidal particles},}\
  }\href@noop {} {\bibfield  {journal} {\bibinfo  {journal} {Phys. Rev. E.}\
  }\textbf {\bibinfo {volume} {54}},\ \bibinfo {pages} {6541} (\bibinfo {year}
  {1996})}\BibitemShut {NoStop}%
\bibitem [{\citenamefont {Kroon}, \citenamefont {Vos},\ and\ \citenamefont
  {Wegdam}(1998)}]{Kroon98}%
  \BibitemOpen
  \bibfield  {author} {\bibinfo {author} {\bibfnamefont {M.}~\bibnamefont
  {Kroon}}, \bibinfo {author} {\bibfnamefont {W.}~\bibnamefont {Vos}}, \ and\
  \bibinfo {author} {\bibfnamefont {G.}~\bibnamefont {Wegdam}},\ }\bibfield
  {title} {\enquote {\bibinfo {title} {Structure and formation of a gel of
  colloidal disks},}\ }\href@noop {} {\bibfield  {journal} {\bibinfo  {journal}
  {Phys. Rev. E.}\ }\textbf {\bibinfo {volume} {57}},\ \bibinfo {pages} {1962}
  (\bibinfo {year} {1998})}\BibitemShut {NoStop}%
\bibitem [{\citenamefont {Avery}\ and\ \citenamefont {Ramsay}(1986)}]{Avery86}%
  \BibitemOpen
  \bibfield  {author} {\bibinfo {author} {\bibfnamefont {R.}~\bibnamefont
  {Avery}}\ and\ \bibinfo {author} {\bibfnamefont {J.}~\bibnamefont {Ramsay}},\
  }\bibfield  {title} {\enquote {\bibinfo {title} {Colloidal properties of
  synthetic hectorite clay dispersions: Ii. light and small angle neutron
  scattering},}\ }\href@noop {} {\bibfield  {journal} {\bibinfo  {journal} {J.
  Colloid Interface Sci.}\ }\textbf {\bibinfo {volume} {109}},\ \bibinfo
  {pages} {448--454} (\bibinfo {year} {1986})}\BibitemShut {NoStop}%
\bibitem [{\citenamefont {Nicolai}\ and\ \citenamefont
  {Cocard}(2000)}]{Nicolai00}%
  \BibitemOpen
  \bibfield  {author} {\bibinfo {author} {\bibfnamefont {T.}~\bibnamefont
  {Nicolai}}\ and\ \bibinfo {author} {\bibfnamefont {S.}~\bibnamefont
  {Cocard}},\ }\bibfield  {title} {\enquote {\bibinfo {title} {Light scattering
  study of the dispersion of laponite},}\ }\href@noop {} {\bibfield  {journal}
  {\bibinfo  {journal} {Langmuir}\ }\textbf {\bibinfo {volume} {16}},\ \bibinfo
  {pages} {8189--8193} (\bibinfo {year} {2000})}\BibitemShut {NoStop}%
\bibitem [{\citenamefont {Au}\ and\ \citenamefont {Leong}(2015)}]{au15}%
  \BibitemOpen
  \bibfield  {author} {\bibinfo {author} {\bibfnamefont {P.-I.}\ \bibnamefont
  {Au}}\ and\ \bibinfo {author} {\bibfnamefont {Y.-K.}\ \bibnamefont {Leong}},\
  }\bibfield  {title} {\enquote {\bibinfo {title} {Surface chemistry and
  rheology of laponite dispersions - zeta potential, yield stress, ageing,
  fractal dimension and pyrophosphate},}\ }\href@noop {} {\bibfield  {journal}
  {\bibinfo  {journal} {Appl. Clay Sci.}\ }\textbf {\bibinfo {volume} {107}},\
  \bibinfo {pages} {36--45} (\bibinfo {year} {2015})}\BibitemShut {NoStop}%
\bibitem [{\citenamefont {Lin}\ \emph {et~al.}(2019)\citenamefont {Lin},
  \citenamefont {Zhu}, \citenamefont {Wang}, \citenamefont {Chen},
  \citenamefont {Phan-Thien},\ and\ \citenamefont {Pan}}]{lin19}%
  \BibitemOpen
  \bibfield  {author} {\bibinfo {author} {\bibfnamefont {Y.}~\bibnamefont
  {Lin}}, \bibinfo {author} {\bibfnamefont {H.}~\bibnamefont {Zhu}}, \bibinfo
  {author} {\bibfnamefont {W.}~\bibnamefont {Wang}}, \bibinfo {author}
  {\bibfnamefont {J.}~\bibnamefont {Chen}}, \bibinfo {author} {\bibfnamefont
  {N.}~\bibnamefont {Phan-Thien}}, \ and\ \bibinfo {author} {\bibfnamefont
  {D.}~\bibnamefont {Pan}},\ }\bibfield  {title} {\enquote {\bibinfo {title}
  {Rheological behavior for laponite and bentonite suspensions in shear
  flow},}\ }\href@noop {} {\bibfield  {journal} {\bibinfo  {journal} {AIP
  Adv.}\ }\textbf {\bibinfo {volume} {9}},\ \bibinfo {pages} {125233} (\bibinfo
  {year} {2019})}\BibitemShut {NoStop}%
\bibitem [{\citenamefont {Zhu}\ and\ \citenamefont {Granick}(2001)}]{zhu01}%
  \BibitemOpen
  \bibfield  {author} {\bibinfo {author} {\bibfnamefont {Y.}~\bibnamefont
  {Zhu}}\ and\ \bibinfo {author} {\bibfnamefont {S.}~\bibnamefont {Granick}},\
  }\bibfield  {title} {\enquote {\bibinfo {title} {Rate-dependent slip of
  newtonian liquid at smooth surfaces},}\ }\href@noop {} {\bibfield  {journal}
  {\bibinfo  {journal} {Phys. Rev. Lett.}\ }\textbf {\bibinfo {volume} {87}},\
  \bibinfo {pages} {096105} (\bibinfo {year} {2001})}\BibitemShut {NoStop}%
\bibitem [{\citenamefont {Joshi}\ and\ \citenamefont
  {Petekidis}(2018)}]{joshi18b}%
  \BibitemOpen
  \bibfield  {author} {\bibinfo {author} {\bibfnamefont {Y.~M.}\ \bibnamefont
  {Joshi}}\ and\ \bibinfo {author} {\bibfnamefont {G.}~\bibnamefont
  {Petekidis}},\ }\bibfield  {title} {\enquote {\bibinfo {title} {Yield stress
  fluids and ageing},}\ }\href@noop {} {\bibfield  {journal} {\bibinfo
  {journal} {Rheol. Acta}\ }\textbf {\bibinfo {volume} {57}},\ \bibinfo {pages}
  {521--549} (\bibinfo {year} {2018})}\BibitemShut {NoStop}%
\end{thebibliography}%

\end{document}